\documentclass[%
 reprint,
 amsmath,amssymb,
 aps,
]{revtex4-1}
\usepackage{bbold}
\usepackage{graphicx}
\usepackage{dcolumn}
\usepackage{bm}

\begin{document}

\preprint{APS/123-QED}

\title{The $\mathcal{PT}$-symmetry-breaking transition in a chain of trapped interacting ions}

\author{Zhenxin Hu}
\affiliation{Guangdong Provincial Key Laboratory of Quantum Metrology and Sensing, and School of Physics and Astronomy, Sun Yat-Sen University (Zhuhai Campus), Zhuhai 519082, China}
\affiliation{State Key Laboratory of Optoelectronic Materials and Technologies, Sun Yat-Sen University (Guangzhou Campus), Guangzhou 510275, China}

\author{Zhenhua Yu}%
 \email{huazhenyu2000@gmail.com}
\affiliation{Guangdong Provincial Key Laboratory of Quantum Metrology and Sensing, and School of Physics and Astronomy, Sun Yat-Sen University (Zhuhai Campus), Zhuhai 519082, China}
\affiliation{State Key Laboratory of Optoelectronic Materials and Technologies, Sun Yat-Sen University (Guangzhou Campus), Guangzhou 510275, China}%


%

\date{\today}

\begin{abstract}
Trapped ions are an ideal platform to implement quantum simulation. Previously the parity-time reversal ($\mathcal{PT}$) symmetry-breaking transition in the paradigmatic non-Hermitian Hamiltonian $h_{PT}=J\sigma_x-i\Gamma\sigma_z$ has been observed in a single ion experiment in a passive way. In this work, we propose to study the interaction effects on the $\mathcal{PT}$-symmetry-breaking transition in a chain of $N$ trapped interacting ions. We consider an effective Ising interaction $H^\text{Ising-x}_\text{int} =\sum_{j<k}U_{jk}\sigma_x^j\sigma_x^k$ between the ions on top of $h_{PT}$. We 
find that sufficiently strong interaction strength can enhance the $\mathcal{PT}$-symmetric phase for even $N$ while the phase is suppressed in all the other cases. In particular, the suppression can be so strong that even infinitesimal dissipation, quantified by $\Gamma$, can turn the system into the $\mathcal{PT}$-symmetry-breaking phase.
In addition, we assess the convolved effects due to the coupling and the spin phase shifts. Our findings can be readily tested in ion chain experiments.
\end{abstract}

\maketitle

\section{\label{sec:1}Introduction}
Non-Hermitian Hamiltonians endowed with the parity-time-reversal ($\mathcal{PT}$) symmetry 
can maintain \emph{real} eigenvalues in certain parameter regimes \cite{Bender1998,Bender2007}. 
Consequently, an unconventional phase transition can occur in such a non-Hermitian system at a kind of singular points called the exceptional points (EP), across which a number of the Hamiltonian eigenvalues switch between being real and complex \cite{Heiss_2004, Heiss_2012}; the transition separates the $\mathcal{PT}$-symmetric phase and the $\mathcal{PT}$-symmetry-breaking phase. An illustrative example is the non-Hermitian Hamiltonian $h_{PT}=J\sigma_x-i\Gamma \sigma_z$, where $J$ is the Rabi frequency and $\Gamma$ quantifies the strength of dissipation \cite{Swanson2004}. This model has the $\mathcal{PT}$-symmetry since $\sigma_x h^*_{PT}\sigma_x=h_{PT}$. Correspondingly, the $\mathcal{PT}$-symmetry-breaking transition occurs at $\Gamma_{\rm tr}=J$. 

Time evolution under non-Hermitian Hamiltonians with the $\mathcal{PT}$-symmetry is expected to be significantly different in the $\mathcal{PT}$-symmetric and breaking phases, which may be harnessed to realize high sensitivity to external disturbances near the EPs, and find potential applications in metrology and sensing \cite{Wiersig2014, Hodaei2017}. Furthermore, such non-Hermitian systems have been theoretically shown to exhibit intriguing phenomena such as unidirectional invisibility \cite{Lin2011}, the non-Hermitian skin effect \cite{Yao2018}, and the non-Bloch oscillations \cite{Longhi2019}. Experimentally, in addition to electromagnetic \cite{Guo,Kip, Kottos, Peschel, Schafer, Yang}, and mechanical systems \cite{Bender2013}, the model $h_{PT}$ has been realized in  single nitrogen-vacancy center in diamond \cite{Du2019}, noninteracting Fermi gases \cite{Luo2019}, and a single trapped ion \cite{Zhang2021}. 

At the same time, theoretical studies are gradually revealing interaction effects on the $\mathcal{PT}$-symmtry-breaking transition \cite{Yu2019, Cui2019, Chen2020, Huang2020,Lourenco2022, Pan2023}. For example, in the context of interacting bosons loaded in a double-well potential, Ref.~\cite{Yu2019} introduced, in addition to the non-Hermitian single particle Hamiltonian $h_{PT}$, an interacting Hamiltonian $H^\text{Ising-z}_\text{int}=g\sum_{j<k}\sigma_z^j\sigma_z^k$ with the superscript indexing different particles. There $\Gamma_\text{tr}/J$ is found to decrease gradually in a monotonic way from its noninteracting value, unity, to zero as $|g|$ increases; the interaction suppresses the $\mathcal{PT}$-symmetric phase as weaker dissipation, i.e., $\Gamma_\text{tr}<J$, is capable of bringing about the transition. Furthermore, when $|g|$ is sufficiently large, for $N$ bosons, it is shown that $\Gamma_{\rm tr}/J\sim|g/J|^{-(N-1)}$; the exponential suppression of $\Gamma_{\rm tr}$ with respect to $N$ indicates that in experiments where $N$ is considerably substantial, it will be rather challenging to locate the transition in the presence of the interaction since the $\mathcal{PT}$-symmetric phase is exponentially suppressed and turning on even a rather small $\Gamma$ is highly likely to put the system directly in the $\mathcal{PT}$-symmetry-breaking phase. Similar suppression is also found when considering the $xy$-interaction $H^\text{xy}_\text{int}=\sum_{j<k}V_{j,k}(\sigma_x^j\sigma_x^k+\sigma_y^j\sigma_y^k)$ potentially realizable in Rydberg atom arrays \cite{Lourenco2022}.
However, how generalizable are the conclusions obtained so far with regard to other forms of interactions? To answer the quenstion, further theoretical studies and experimental tests are needed. 

Experimentally, trapped ions are an ideal platform to simulate interaction effects of non-Hermitian Hamiltonians with the $\mathcal{PT}$-symmetry \cite{Monroe2021}. Manipulation of trapped constituent ions in present day labs is remarkably sophisticated \cite{Monroe2021, Wineland1992}. Effective spin interactions have been realized among an ensemble of ions \cite{Monroe2021, Cirac2004}. Combining the capability of realizing $h_{PT}$ for each single ions \cite{Zhang2021} and effective interactions between the ions, simulation in such systems holds the prospect to enrich our knowledge on how interactions affect properties of non-Hermitian systems. 

In this work, we study a chain of $N$ ions, in which the ions interact with an Ising coupling $H^\text{Ising-x}_\text{int} =\sum_{j<k}U_{jk}\sigma_x^j\sigma_x^k$ \cite{Monroe2021}, while the single ion Hamiltonian is given by $h_{PT}$. It has been shown that interactions can change the order of exceptional points \cite{Cui2019b,Zhang2022}. Here we focus on the interaction effects on the $\mathcal{PT}$-symmetry-breaking transition. We find that when $|U|$, the overall magnitude of $U_{jk}$, is tuned, the $\mathcal{PT}$-symmetric phase can be either suppressed or enhanced, depending on the values of $N$ and $|U|$. In particular, there are finite number of zeros of $\Gamma_\text{tr}$ occurring at certain values of $|U|/J$; right at these zeros, even infinitesimally weak dissipation is capable of rendering the system in the $\mathcal{PT}$-symmetry-breaking phase. 
On the other hand, when $|U|/J\gg1$, the transition is found to follow the scaling $\Gamma_{\rm tr}\sim |U|^0$ for odd $N$ and $\Gamma_{\rm tr}\sim |U|^1$ for even $N$. Therefore there is an interesting disparity between $N$ being odd and even 
that sufficiently large $|U|$ can enhance the $\mathcal{PT}$-symmetric phase for the latter ($\Gamma_{\rm tr}/J>1$) while the phase is generally suppressed in else cases ($\Gamma_{\rm tr}/J<1$). 
We further assess the effects of spin phase shifts that can be introduced in experiments \cite{Monroe2021}, and show that nonzero spin phase shifts can lift the zeros of $\Gamma_{\rm tr}$.
Our results are ready to be tested in experiments.
Comparing our results to previous studies can shed a new light on the rich behaviors of $\mathcal{PT}$-symmetric non-Hermitian systems in the presence of interactions.

\section{\label{sec:2}The $\mathcal{PT}$-symmetric non-Hermitian Hamiltonian}

\begin{figure}[t]
\includegraphics[width=0.45\textwidth]{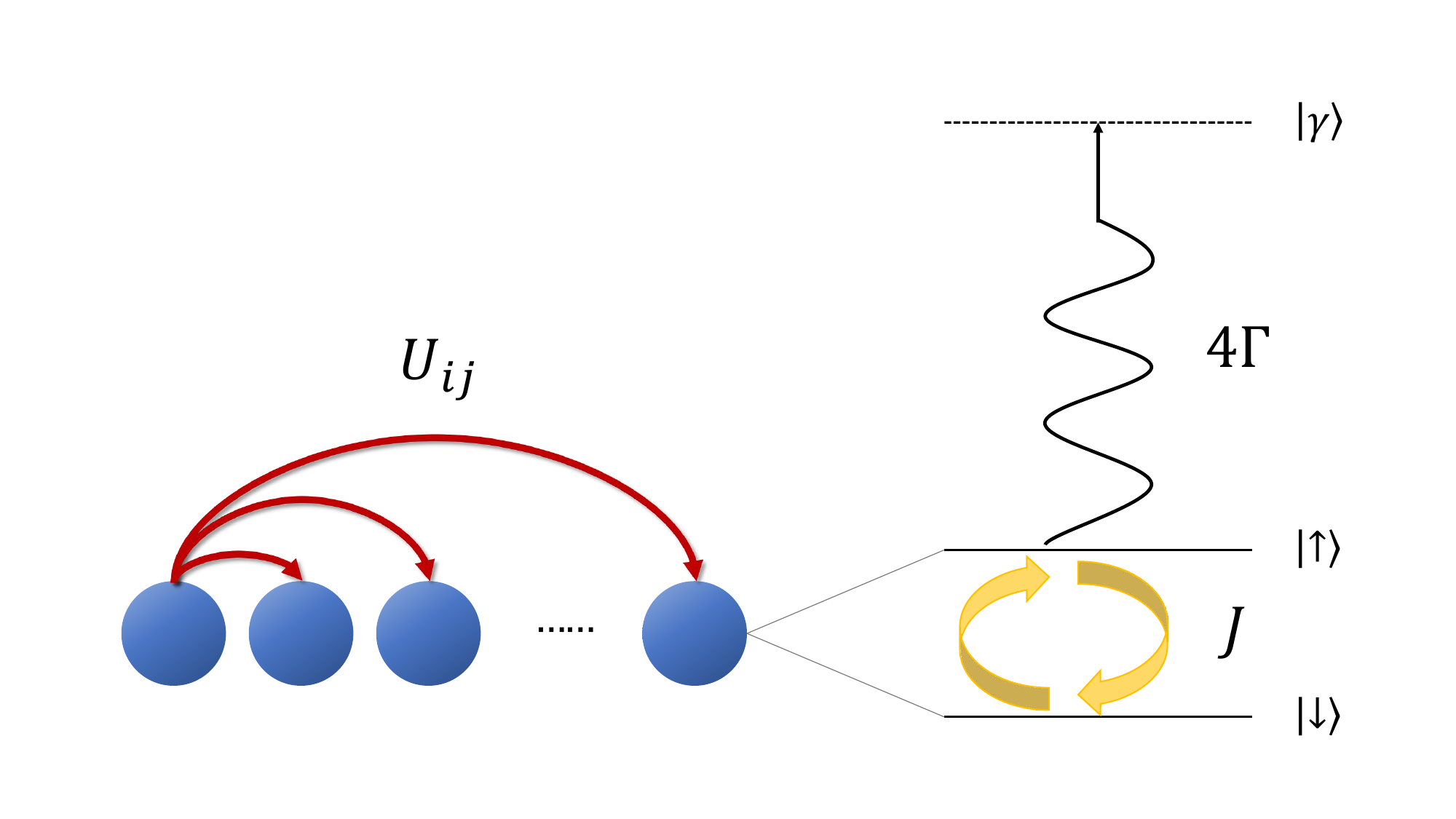}
\caption{\label{fig:1} Schematic of the ion chain. In a chain of $N$ ions, each ion encodes a pseudo-spin via two internal states labeled as $\left|\uparrow \right\rangle$ and $\left|\downarrow \right\rangle$. Raman beams are employed to realize an effective Ising coupling; the coupling strength between the $i$th and $j$th ions is denoted by $U_{ij}$. 
A rf field is used to directly couple the two internal states of each ion with the Rabi frequency $J$.
An additional optical field couples $\left|\uparrow \right\rangle$ to an excited state $\left|\gamma \right\rangle$ and induces population loss from the state $\left|\uparrow \right\rangle$ at rate $4\Gamma$. \label{schematic}}
\end{figure}

Our theoretical study corresponds to  the experimental setup of a chain of ions trapped in a linear radio-frequency (RF) trap \cite{Wineland1992}. The schematic is given in Fig.~\ref{schematic}.
Due to the trap and the inter-ion Coulomb repulsive interactions, the ions form a crystal. Previously, by applying an external microwave field to couple two electronic ground states of $^{171}$Yb$^+$ denoted by $\left| \downarrow\right\rangle$ and $\left| \uparrow\right\rangle$ and a weak optical beam to induce transition between $\left| \uparrow\right\rangle$ and excited electronic states, the non-Hermitian Hamiltonian of a single ion 
\begin{align}
h=J\sigma_x-i2\Gamma \left| \uparrow\right\rangle\left\langle\uparrow\right|\label{h}
\end{align}
has been realized experimentally, where the loss rate of the number of ions in the $\left| \uparrow\right\rangle$ state is $4\Gamma$ \cite{Zhang2021}. Via $h$ in Eq.~(\ref{h}), the experimentalists have successfully simulated the $\mathcal PT$-symmetry-breaking transition of the non-Hermitian Hamiltonian 
\begin{align}
h_{PT}=J\sigma_x-i\Gamma \sigma_z\label{hpt}
\end{align}
in ions \emph{passively} since $h_{PT}=h+i\Gamma\mathbb{1}$ \cite{Zhang2021}.

On the other hand, the Raman process involving the states $\left|\downarrow\right\rangle$ and $\left|\uparrow\right\rangle$ 
can result in an effective Ising spin interaction between the ions \cite{Monroe2021}
\begin{align}
H^\text{Ising-x}_\text{int}=\sum_{j<k}U_{jk}\sigma_x^j\sigma_x^k,\label{hint}
\end{align}
where the subscripts of $U_{jk}$ and the superscript of $\boldsymbol \sigma^j$ label the ions along the chain. This interaction is of the Ising coupling form.
The effective spin interaction coupling $U_{jk}$ is well approximated by an inverse power form with distance between the ions, i.e., $U_{jk}=U/{\left|j-k\right|}^\alpha$ \cite{Monroe2021}; the nearest-neighbour coupling is $U$ and the exponent $\alpha$ can be experimentally tuned from $0$ to $3$ by changing the Raman laser frequencies \cite{Cirac2004}. 

By combining the passive realization of $h_{PT}$ in Eq.~(\ref{hpt}) and the effective spin interaction Hamiltonian $H^\text{Ising-x}_\text{int}$ in Eq.~(\ref{hint}), one can in experiments simulate the resulting 
effective non-Hermitian Hamiltonian of the interacting ion chain
\begin{align}
H&=H_0+H^\text{Ising-x}_\text{int}\nonumber \\
H_0&=\sum_{j=1}^{N}h_{PT}^j=2JS_x-2i\Gamma S_z\label{htot},
\end{align}
with $N$ the total number of the ions, and the total pseudo-spin operators $\mathbf S\equiv\sum_{j=1}^{N}\boldsymbol\sigma^j/2$
. It is obvious that $H^\text{Ising-x}_\text{int}$ maintains the $\mathcal{PT}$-symmetry since $(\prod_j\sigma_x^j)(H^\text{Ising-x}_\text{int})^*(\prod_k\sigma_x^k)=H^\text{Ising-x}_\text{int}$. Our task is to theoretically reveal the effect of $H^\text{Ising-x}_\text{int}$ on the 
$\mathcal{PT}$-symmetry-breaking transition.

\section{\label{sec:3}The $\mathcal{PT}$-symmetry-breaking transition}

The $\mathcal{PT}$-symmetry-breaking transition is known to associate with the coalesce of
eigenvalues of $\mathcal{PT}$-symmetric non-Hermitian Hamiltonians \cite{Alu2021}. The addition of $H^\text{Ising-x}_\text{int}$ to $\sum_{j=1}^{N}h_{PT}^j$ modifies the eigenvalues of the ion chain and thus is expected to affect the transition. In the following, we begin our investigation with the case $\alpha=0$ for $U_{jk}=U/{\left|j-k\right|}^\alpha$. This infinite ranged case allows us to understand the interaction effects analytically, and further provides basis to understand the numerical results obtained for finite ranged cases.

\subsection{\label{sec:level2}The infinite ranged Ising coupling}

When $\alpha=0$, the non-Hermitian Hamiltonian, Eq.~(\ref{htot}), becomes
\begin{align}
H=2JS_x-2i\Gamma S_z+2U\left(S_x^2-\frac{N}{4}\right)\label{hai}.
\end{align}
Suppose experimentally we start with the initial state $\left|\chi(t=0)\right\rangle$ that all the spins are in the state $\left|\downarrow\right\rangle$. Since $S_z\left|\chi(t=0)\right\rangle=N/2$ and $[\mathbf{S},H]=0$, $\left|\chi(t)\right\rangle$ is confined in the Hilbert subspace of $S=N/2$. Thus we are concerned with only the eigenvalues of $H$ in this subspace.

We choose the basis of the subspace as $\left|N/2,m\right\rangle$ with $m= -N/2,-N/2+1,\cdots ,N/2-1,N/2$,
where $S_x\left|N/2,m\right\rangle=m\left|N/2,m\right\rangle$, $\mathbf{S}^2\left|N/2,m\right\rangle=N(N+2)/4\left|N/2,m\right\rangle$. This basis diagonalizes the interaction part of $H$.
The matrix element of the total Hamiltonian is given by 

\begin{align}
H_{m'm}  =&  \left \langle N/2,m' \right | H\left | N/2,m  \right \rangle \notag\\
 = & \left [ 2J m+2U\left( m^2 -N/4\right ) \right ]  \delta_{m'm}\notag\\
&   -  i\Gamma\sqrt{(N/2)(N/2+1)-m^2-m} \delta_{m',m+1}\notag\\
&   -  i\Gamma\sqrt{(N/2)(N/2+1)-m^2+m} \delta_{m',m-1}
\label{hm}.
\end{align}
By diagonalizing the matrix of the Hamiltonian, we can obtain its eigenvalues. 

\begin{figure}[t]
\includegraphics[width=0.50\textwidth]{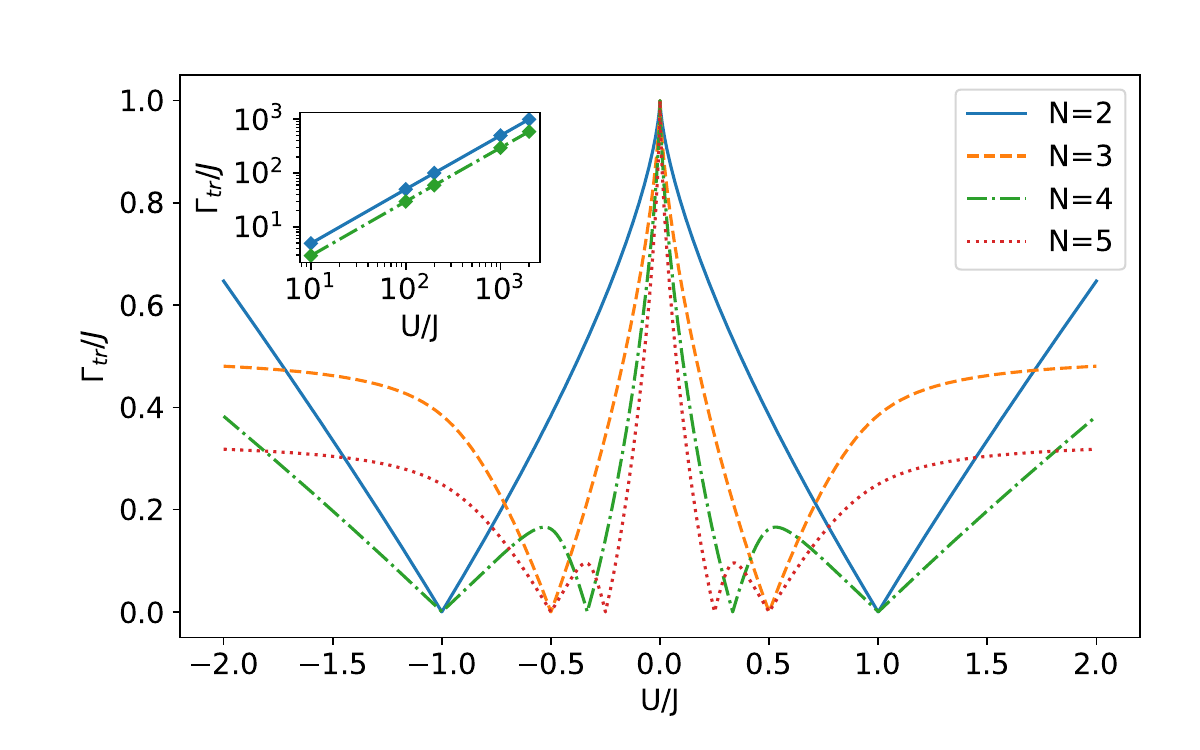}
\caption{\label{ai} Critical value $\Gamma_{\rm tr}$ at the $\mathcal{PT}$-symmetry-breaking transition versus  $U$ for the number of ions $N=2,3,4,5$ derived from Eq.~(\ref{hai}). Below the transition lines are the $\mathcal {PT}$-symmetric phases, and above are the $\mathcal {PT}$-symmetry-breaking phases. The inset shows the asymptotes of $\Gamma_{\rm tr}$ when $U$ is large for $N=2,4$.}
\end{figure}

When $\Gamma$ is sufficiently small, the system is in the $\mathcal{PT}$-symmetric phase where all the eigenvalues of the non-Hermitian Hamiltonian are real. As $\Gamma$ increases, these eigenvalues keep moving on the real axis. At the transition point $\Gamma=\Gamma_{\rm tr}$, some of the eigenvalues coalesce. When $\Gamma$ exceeds $\Gamma_{\rm tr}$, the system enters into the $\mathcal{PT}$-symmetry-breaking phase, and the coalescing eigenvalues split again and acquire imaginary parts \cite{Alu2021}. Figure \ref{ai} plots the transition value $\Gamma_{\rm tr}$ versus the interaction strength $U$ for various numbers of ions $N$. It is clear from Eq.~(\ref{htot}) that since the Hamiltonian $H(U)$ is related to $-H(-U)$ by a unitary transformation, $\Gamma_{\rm tr}$ is an even function of $U$. When $U$ is small, $\Gamma_{\rm tr}/J$ is found to decrease from unity with $|U|$, which can be easily confirmed by a perturbation calculation \cite{Yu2019}. In the large $|U|$ limit, $\Gamma_{\rm tr}/J$ grows as $\sim |U|$ for even number of ions, and saturates to a constant value for odd number of ions. Between the small and large $|U|$ limits, $\Gamma_{\rm tr}/J$ has $N$ zeros for even $N (\ge2)$ and $N-1$ zeros for odd $N (\ge3)$. Thus, compared to $\Gamma_\text{tr}/J=1$ at $|U|=0$, the interaction is found to enhance the $\mathcal{PT}$-symmetric phase for even $N$ ($\Gamma_\text{tr}/J>1$) when $|U|$ is large enough, while the phase is generally suppressed by the interaction in all the other cases ($\Gamma_\text{tr}/J<1$).

The emergence of zeros of $\Gamma_{\rm tr}$ is because in the absence of dissipation, i.e., $\Gamma=0$, not only $[H, S_x]=0$ but also $[H^\text{Ising-x}_\text{int},S_x]=0$. Such that when $U$ is tuned, $\{|N/2,m\rangle\}$ keep being the eigen-states of $H$ and their eigen-values, $\lambda_m=2Jm+2U(m^2-N/4)$, which change with $U$, can \emph{cross} (in contrast to anticrossing). Actually two adjacent eigenvalues $\lambda_m$ and $\lambda_{m-1}$ become degenerate whenever $J/U=1-2m$ (note $J$ is nonzero, $m$ cannot take $1/2$). At any one of the degenerate points, turning on an infinitesimal dissipation, i.e., including $-2i\Gamma S_z$ in $H$, would give rise to (infinitesimal) imaginary parts to the two originally degenerate eigenvalues since $\langle N/2,m|S_z|N/2,m-1\rangle\neq0$, and end up the ions in the $\mathcal{PT}$-symmetry-breaking phase. All these zeros of $\Gamma_{\rm tr}$ are thus confined in the region $-1\le U/J\le1$ as shown in Fig.~(\ref{ai}). The existence of zeros of $\Gamma_{\rm tr}$ determines that in their vicinities, the $\mathcal{PT}$-symmetric phase must be suppressed.

\begin{figure}[t]
\includegraphics[width=0.50\textwidth]{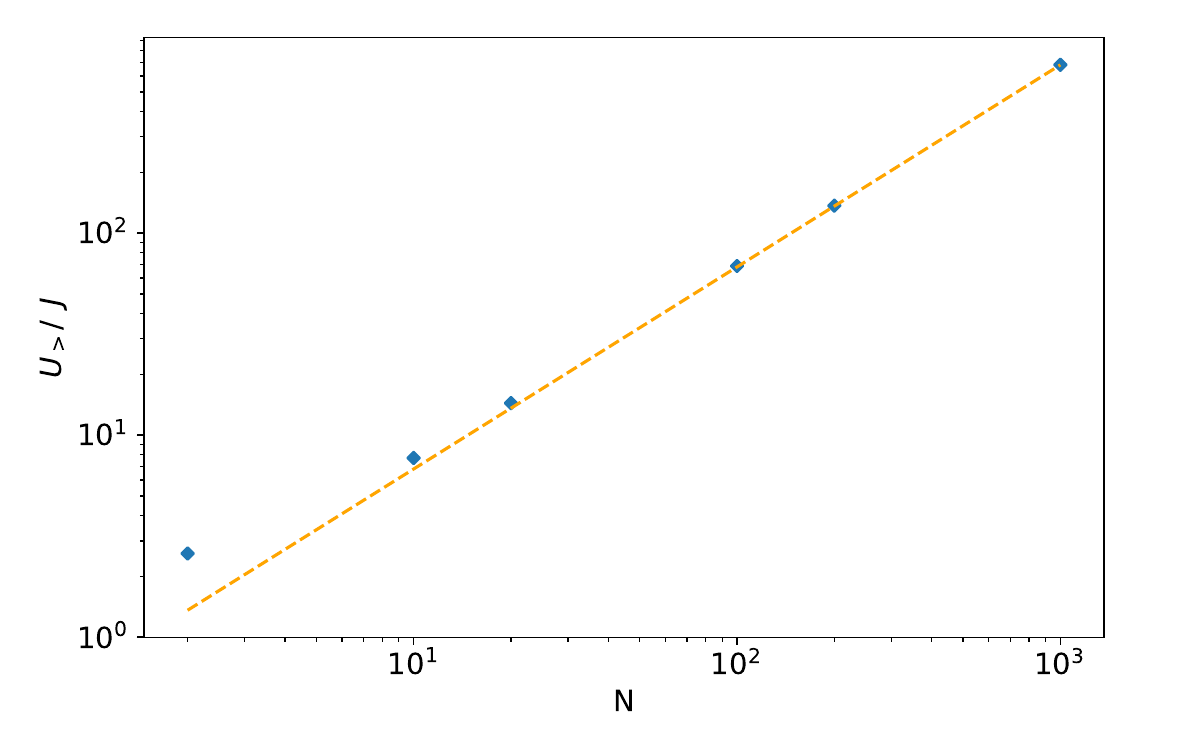}
\caption{\label{ub} The interaction threshold $U_>$ versus $N$. When $U>U_>$, the $\mathcal{PT}$-symmetric phase is enhanced, i.e., $\Gamma_{\rm tr}/J>1$. The dashed line has slope equal to one in the Log-Log plot, and is meant to guide eyes.}
\end{figure}

In another limiting situation $|U|/J\gg1$, for odd $N$, the transition can be attributed to the subspace spanned by $|N/2,1/2\rangle$ and $|N/2,-1/2\rangle$ in which the diagonal interaction energies $\sim U$ are the same; coalescing of eigenenergies of $H$ in the subspace requires the discriminant $\Delta= \left [ J^2-(N+1)^2\Gamma^2/4\right ] =0$, which yields  $\Gamma_{\rm tr}\to 2J/ (N+1)$, agreeing with Fig.~\ref{ai}. On the other hand, for even $N$, we can likewise truncate the 
Hamiltonian matrix into the subspace spanned by $|N/2,m\rangle$ with $m=-1,0,1$, 
only two eigenvalues of the truncated Hamiltonian can coalesce, requiring the discriminant $\Delta\approx \left [4U^2-(2N^2+4N)\Gamma^2\right] =0$, yielding $\Gamma_{\rm tr}\to\left|U\right|/\sqrt{N^2/2+N}$. 
Compared to Fig.~\ref{ai}, the approximation of truncation succeeds in explaining the scaling $\Gamma_{\rm tr}\sim c |U|$ for even $N$ in the limit $|U|/J\to\infty$, and the general trend that the prefactor $c$ decreases with larger $N$. The decreasing $c$ with $N$ also indicates that there exists an interaction threshold $U_>$ such that when $U>U_>$, the $\mathcal{PT}$-symmetric phase is enhanced, i.e., $\Gamma_{\rm tr}/J>1$. Figure \ref{ub} shows that $U_>$ scales linearly in $N$ for large $N$, which is compatible with the result $\Gamma_{\rm tr}\to\left|U\right|/\sqrt{N^2/2+N}$ obtained from truncation. (See Appendix for more details).

\subsection{\label{sec:level3}The finite ranged Ising coupling}
\begin{figure*}
	\centering
	\includegraphics[width = 0.95\linewidth]{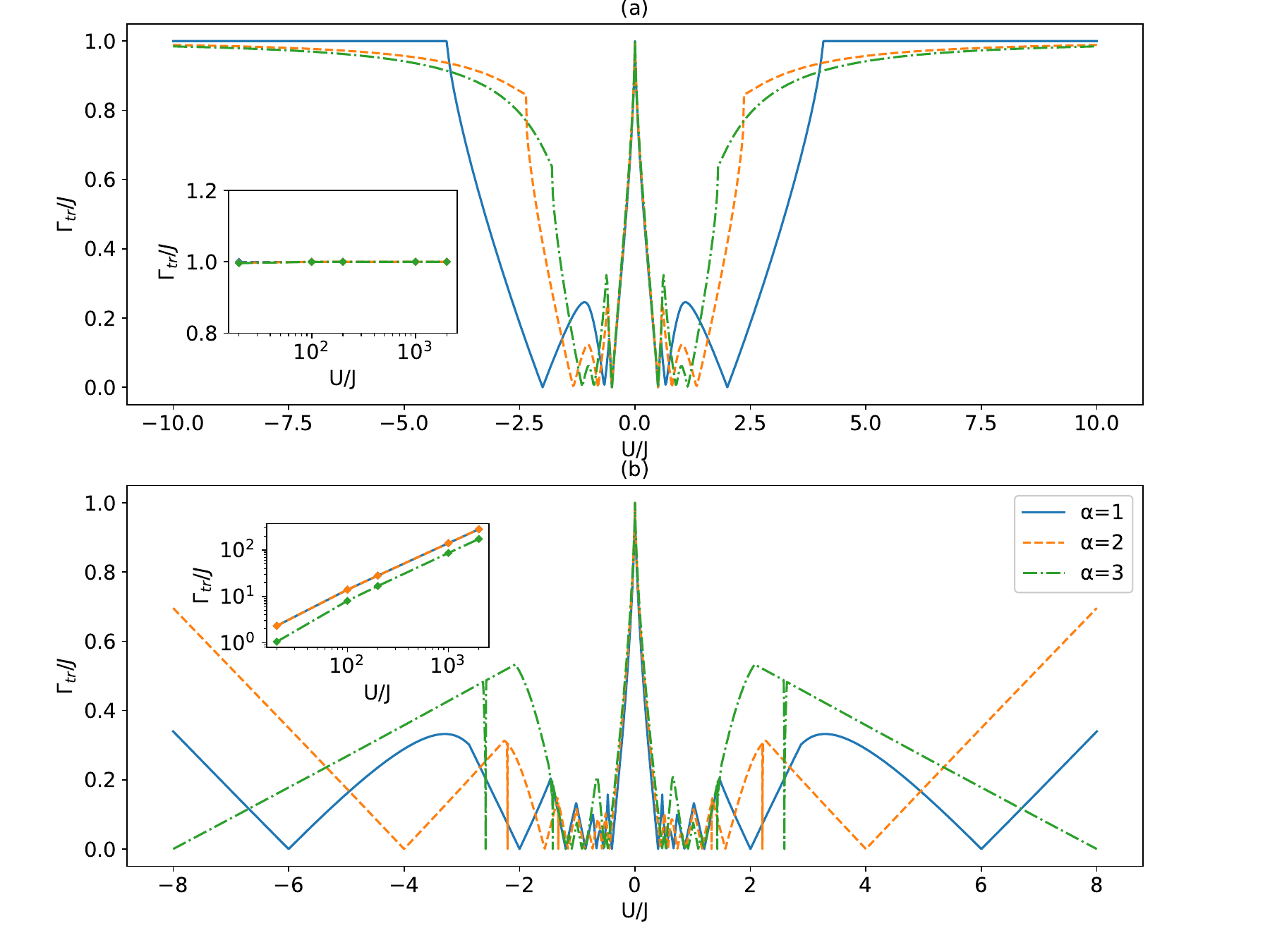}
	\caption{\label{fig:3} Critical value $\Gamma_{\rm tr}$ at the $\mathcal{PT}$-symmetry-breaking transition versus $U$ with $\alpha=1,2,3$ for the number of ions $N=3$ (a) and $N=4$ (b) derived from Eq.~(\ref{htot}). The insets show the asymptotes of $\Gamma_{\rm tr}$ when $U$ is large.}
	\end{figure*}

More generally, the exponent $\alpha$ for the Ising coupling can be modified from $0$ to $3$ by adjusting the optical detuning in experiments~\cite{Cirac2004}. 
When $\alpha\neq0$, even though we start with an initial state in the multiplet $S=N/2$, subsequent evolution of the state could spill over out the multiplet. We have to determine the transition by diagonalizing the full $2^N\times 2^N$ Hamiltonian matrix. Figure~\ref{fig:3} shows the transition boundaries of examplary cases $N=3,4$ for $\alpha=1,2,3$. Compared with Fig.~\ref{ai} for $\alpha=0$, features, such as the existence of zeros of $\Gamma_{\rm tr}$ and the scaling of $\Gamma_{\rm tr}$ in terms of $|U|$ for large $|U|$, and correspondingly the suppression and enhancement of the $\mathcal{PT}$-symmetric phase, are maintained. Though for $\alpha\neq0$ the interaction Hamiltonian $H^\text{Ising-x}_\text{int}=\sum_{j<k}U\sigma_x^j\sigma_x^k/|j-k|^\alpha$ can no longer be casted solely in terms of $S_x$, $\{|S,m\rangle\}$ are still the simultaneous eigen-states of $H^\text{Ising-x}_\text{int}$ and $H$ at $\Gamma=0$. As reasoned above for the case of $\alpha=0$, there are zeros of $\Gamma_{\rm tr}/J$; however, more zeros of $\Gamma_{\rm tr}/J$ are found to appear at less ``regular" values of $U/J$, and the locations of these zeros shift with $\alpha$ and can exceed the region $-1\le U/J\le1$ (see Appendix for details).

\subsection{\label{Sec.3} Effects of spin phase shifts}
\begin{figure}[h]
\includegraphics[width=0.45\textwidth]{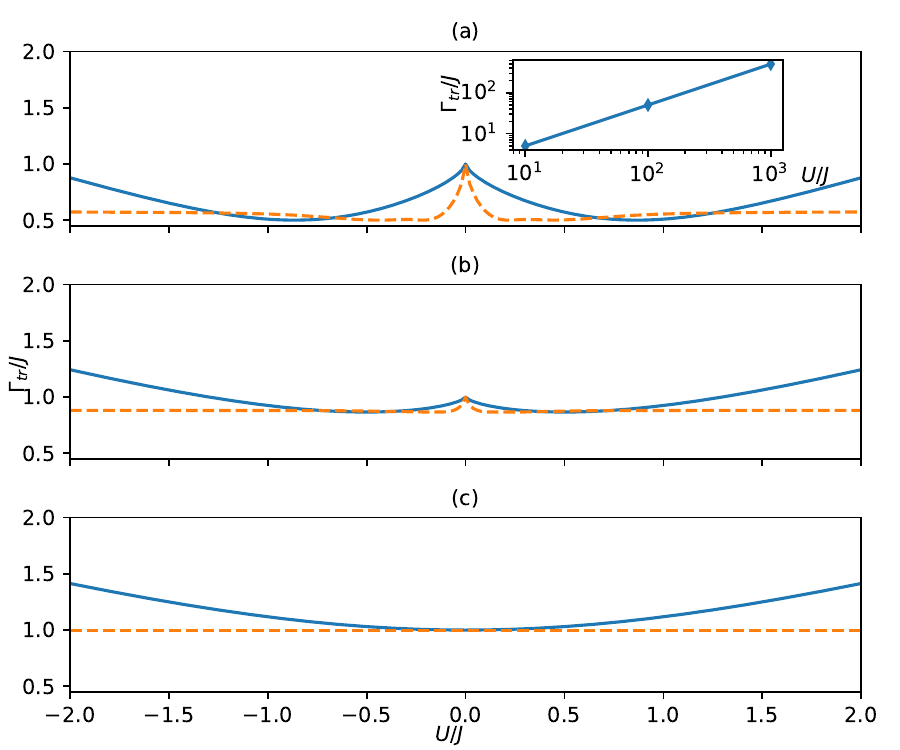}
\caption{\label{fig:4} Critical value $\Gamma_{\rm tr}$ versus $U$ with a uniform spin phase shift $\theta = \pi /4 (a),\pi /3 (b),\pi /2 (c)$ derived from Eq.~(\ref{htheta}). The (blue) solid lines are for $N=2$ and the (orange) dashed lines are for $N=5$. The inset shows the asymptote of $\Gamma_{\rm tr}$ for $N=2$ when $U$ is large.}\label{theta}
\end{figure}

In experiments of trapped ions, implementation to generate the effective Ising interaction Hamiltonian, Eq.~(\ref{hint}), may imprint spin phase shifts \cite{Monroe2021}, i.e., the spin operator of the $j$th ion appearing in Eq.~(\ref{hint}) is $\sigma^j_{\theta_j}=\cos(\theta_j)\sigma^j_x+\sin(\theta_j)\sigma^j_y$ instead of $\sigma^j_x$ ($\theta_j$ can be different for different $j$). To reveal the effects of such spin phase shifts on the transition, we apply the unitary transformation $U_j=e^{-i\theta_j\sigma^j_z/2}$ such that $U_j^\dagger \sigma^j_{\theta_j} U_j=\sigma^j_x$; the transformed Hamiltonian is given by
\begin{align}
H'(\{\theta_j\})&=-i\Gamma\sum_j \sigma^j_z +J\sum_j\left[\cos(\theta_j)\sigma_x^j-\sin(\theta_j)\sigma_y^j\right]\notag\\
&+\sum_{j<k}U_{jk}\sigma_x^j\sigma_x^k.\label{htheta}
\end{align}
It is obvious that $H'$ maintains the $\mathcal{PT}$-symmetry, i.e., $(\prod_j\sigma_x^j)(H')^*(\prod_k\sigma_x^k)=H'$. 
For simplicity, let us consider the uniform situation, $\theta_j=\theta$, independent of $j$. Figure \ref{theta} shows $\Gamma_{\rm tr}$ for examplary cases $N=2, 5$ with $\alpha=0$ and various uniform spin phase shift $\theta=\pi/4,\pi/3,\pi/2$ respectively. Compared with Fig.~\ref{ai}, a prominent difference is the disappearance of the zeros of $\Gamma_{\rm tr}/J$ in Fig.~\ref{theta} as $U/J$ varies. This difference is due to the presence of the term $-J\sum_j \sin(\theta_j)\sigma_y^j$ in $H'$, which is nonzero when $\theta_j\neq0$ for any $j$. This term renders $\{|S,m\rangle\}$ no longer the eigenstates of $H'$ at $\Gamma=0$. Actually for $H'$, there is no operator $O$ such that $[O, H^\text{Ising-x}_\text{int}]=0$ and $[O,H']$ at $\Gamma=0$ except the possibility $O=S^2$ when $\theta_j$ is uniform and $\alpha=0$. Therefore, the mechanism giving rise to the zeros of $\Gamma_\text{tr}$ in previous cases is missing. A straightforward example is $\theta=\pi/2$, whose Hamiltonian is $H'(\{\theta_j=\pi/2\})=-2JS_y-2i\Gamma S_z+2U\left(S_x^2-N/4 \right)$. At $\Gamma=0$, as $U$ changes, the eigen-levels of $H'(\{\theta_j=\pi/2\})$ can only anti-cross each other (instead of crossing). 
In company with the disappearance of the zeros of $\Gamma_{\rm tr}$, the suppression of the $\mathcal{PT}$-symmetric phase can be reduced. For example, the suppression is gone at all in Fig.~\ref{theta}(c) while the enhancement is strengthened. 

\subsection{\label{Sec.4}Experimental signature
}
To determine the $\mathcal{PT}$-symmetry-breaking transition modified by the interaction in experiments, we note that the density matrix $\rho$ of the $N$ interacting ions in the concerned Hilbert space spanned by the internal states $\left|\uparrow \right\rangle$ and $\left|\downarrow \right\rangle$ for each ions shall obey the master equation
\begin{equation}
\frac{ \partial\rho(t)}{\partial t}=-i\left[\mathcal{H},\rho\right]+2\Gamma D\left[\{b_{j,\uparrow}\}\right]\rho(t),
\end{equation}
where $\mathcal{H}$ is the Hermitian part of Eq.~(\ref{htot}), and the dissipator is $D\left[\{b_{j,\uparrow}\}\right]\rho=\sum_j [2b_{j,\uparrow}\rho b^\dagger_{j,\uparrow}-(b_{j,\uparrow}^\dagger b_{j,\uparrow}\rho+\rho b_{j,\uparrow}^\dagger b_{j,\uparrow})]$ with $b_{j,\uparrow}$ and $b_{j,\uparrow}^\dagger$ being the annihilation and creation operators of the $j$th ion in the $\left|\uparrow \right\rangle$ internal state [the sum of $\mathcal{H}$ and $-i2\Gamma b_\uparrow^\dagger b_\uparrow$ from the dissipator equals that of $H$ from Eq.~(\ref{htot}) and $-i\Gamma(b_\uparrow^\dagger b_\uparrow+b_\downarrow^\dagger b_\downarrow)$].

\begin{figure}[t]
\includegraphics[width=0.45\textwidth]{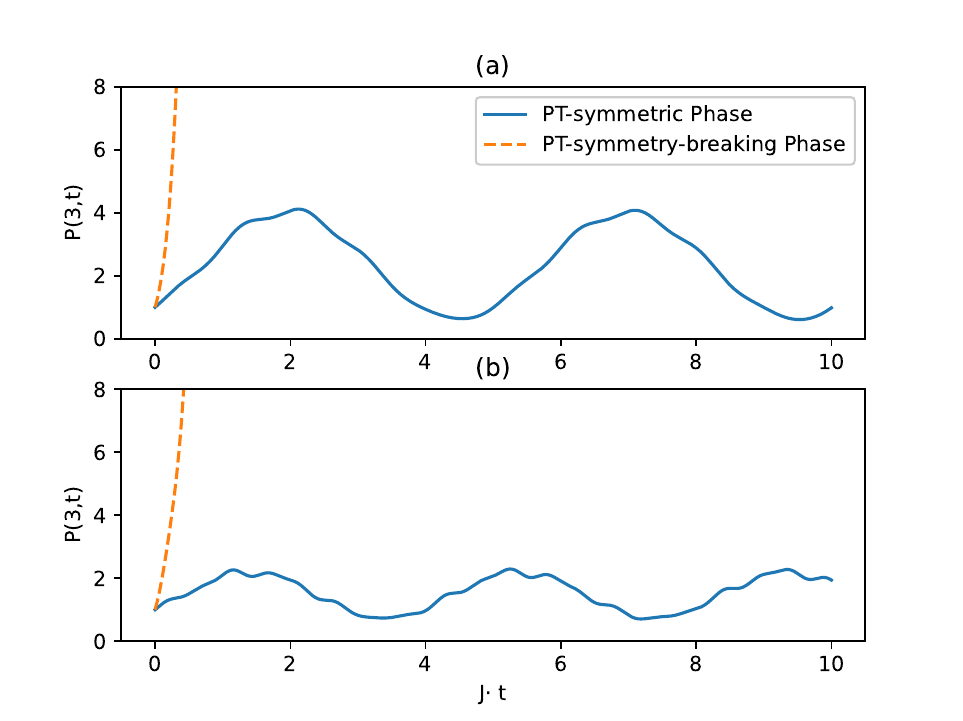}
\caption{\label{fig:5} Time evolution of $P(3,t)$. (a) $\alpha=0$. With $U/J=1, \Gamma/J=1.2$, the system is in the $\mathcal PT$-symmetry-breaking phase, and $P(3,t)$ grows exponentially. With $U/J=1, \Gamma/J=0.3$, the system is in the $\mathcal PT$-symmetric phase, and $P(3,t)$ oscillates and is bounded. (b) $\alpha=2$. $P(3,t)$ is plotted in the $\mathcal PT$-symmetric phase with $U/J=2.5, \Gamma/J=3.5$ and in the $\mathcal PT$-symmetric phase with $U/J=2.5, \Gamma/J=0.75$.}
\end{figure}

In experiments, suppose that we start with $N$ ions prepared in the internal states spanned by $\left|\uparrow \right\rangle$ and $\left|\downarrow \right\rangle$, the density matrix $\rho_{NN}(t)$ in the subspace spanned by these internal states is given by $\rho_{NN}(t)=e^{-N\Gamma t}e^{-iHt}\rho_{NN}(0)e^{iH^\dagger t}$. Thus the $\mathcal{PT}$-symmetry-breaking transition shall be reflected in the properties of the quantity $e^{N\Gamma t}\rho_{NN}(t)$. 
Assuming that one can detect the number of ions remaining in the internal states spanned by $\left|\uparrow \right\rangle$ and $\left|\downarrow \right\rangle$ with high accuracy in experiments, the time evolution of the rescaled probability $P(N,t) \equiv e^{N\Gamma t}{\rm Tr}\rho_{NN}(t)={\rm Tr}\left[e^{-iH_{PT}t}\rho_{NN}(0)e^{iH_{PT}^\dagger t}\right]$ can be used as the observable to distinguish the symmetric and symmetry-breaking phases. Of course, alternative post-selection measurement may be employed \cite{Lourenco2022}. Figure (\ref{fig:5}) plots the time evolution of $P(3,t)$ with the initial state $|\downarrow\downarrow\downarrow\rangle$ in the two phases for $\alpha=0,2$; $P(3,t)$ oscillates in the $\mathcal{PT}$-symmetric phase and diverges in the breaking phase. Thus in experiments, one can determine the transition by distinguishing the qualitatively different time evolution behavior of $P(N,t)$. 

\section{\label{Sec.4}Discussion}
We have studied the $\mathcal{PT}$-symmetry-breaking transition in a chain of trapped interacting ions. We have shown how the Ising coupling between the ions can change the critical value of $\Gamma_{\rm tr}$, making it different from the one determined by $h_{PT}$. Remarkable features include that the Ising coupling, when fine tuned, can suppress $\Gamma_{\rm tr}$ to zero. This suppression is due to the presence of the conserved quantity $S_x$ when $\Gamma=0$; the conserved $S_x$ enables crossings of certain eigenstates which can be coupled together via the dissipation $-2i\Gamma S_z$. 
In the limit $|U|/J\gg1$, we find $\Gamma_{\rm tr}\sim |U|^1$ for even $N$ and $\Gamma_{\rm tr}\sim |U|^0$ for odd $N$.
Therefore the $\mathcal{PT}$-symmetric phase can be enhanced for even $N$ when $|U|$ is sufficiently large and suppressed for all the other cases. We show that nonzero spin phase shifts, which can be realized in experiments, can render $S_x$ not conserved any more at $\Gamma=0$, and lift the zeros of $\Gamma_{\rm tr}$. Simultaneously the suppression of the $\mathcal{PT}$-symmetric phase is reduced.

Our finding of $\Gamma_{\rm tr}$ with respect to the interaction strength is different from the work \cite{Yu2019} considering another interaction $H^\text{Ising-z}_\text{int}=g\sum_{j<k}\sigma_z^j\sigma_z^k=2g(S_z^2-N/4)$ where $\Gamma_{\rm tr}$ is found to decrease towards zero monotonically as $|g|$ increases no matter $N$ being even or odd. There the reason for 
the absence of zeros of $\Gamma_\text{tr}$ is essentially the same as when including spin phase shifts in our cases that  
at $\Gamma=0$, there are no operators other than $S^2$ simultaneously commuting with $H^\text{Ising-z}_\text{int}$ and $h_{PT}$ (note $S_x$ does not commute with $H^\text{Ising-z}_\text{int}$). Thus tuning $g$ can only give rise to anti-crossing between the eigen-levels [cf.~crossing between eigen-levels can happen to Eq.~(\ref{hai}) at $\Gamma=0$ when $U$ is tuned], and no degeneracy can arise. Comparison between ours and previous studies concludes that the form of interactions is vital in determining how the critical point $\Gamma_\text{tr}$ is modified by the interactions. 

\section*{Acknowledgements}
This work is supported by the National Natural Science Foundation of China (Grant No.~12074440).

\section*{Appendix}
\begin{center}
\begin{table}
\caption{Degenerate pairs of eigen-energies $\lambda_y$ of Eq.~(\ref{htot}) of $N=3,4$ and $\Gamma=0$ at different values of $U/J$. The eigen-states corresponding to the pairs of $\{\lambda_y, \lambda_{y'}\}$ have the property $\langle y|\sum_j\sigma_z^j|y'\rangle\neq0$.}
\begin{tabular}{|c | c |}
\hline
$U/J$ & Degenerate pairs of $\lambda_y$  for $N=3$\\
\hline
$-1/2$ & \{$\lambda_1$, $\lambda_3$\}\\
$1/2$ & \{$\lambda_6,\lambda_8$\}\\
$-(1+1/2^\alpha)^{-1}$ & \{$\lambda_1$, $\lambda_2$\}, \{$\lambda_1$, $\lambda_5$\}\\
$-(1-1/2^\alpha)^{-1}$ & \{$\lambda_2$, $\lambda_6$\}, \{$\lambda_5$, $\lambda_6$\}\\
$(1-1/2^\alpha)^{-1}$ & \{$\lambda_3$,  $\lambda_4$\}, \{$\lambda_3$, $\lambda_7$\}\\
$(1+1/2^\alpha)^{-1}$ & \{$\lambda_4$, $\lambda_8$\}, \{$\lambda_7$, $\lambda_8$\}\\
\hline
\end{tabular}
\hspace{200px}
\begin{tabular}{|c | c |}
\hline
$U/J$ & Degenerate pairs of $\lambda_y$ for $N=4$\\
\hline
$-(1+2^{-\alpha}+3^{-\alpha})^{-1}$ & \{$\lambda_1$, $\lambda_2$\}\\
$(1+2^{-\alpha}+3^{-\alpha})^{-1}$ & \{$\lambda_{15}$, $\lambda_{16}$\}\\
$-3(1+2^{-\alpha}+3^{-\alpha})^{-1}$ &\{$\lambda_1$, $\lambda_8$\}\\
$3(1+2^{-\alpha}+3^{-\alpha})^{-1}$ & \{$\lambda_{9}$, $\lambda_{16}$\}\\
$-(2+2^{-\alpha})^{-1}$ & \{$\lambda_1$, $\lambda_5$\}\\
$(2+2^{-\alpha})^{-1}$ & \{$\lambda_{12}$, $\lambda_{16}$\}\\
$-(1+2^{-\alpha}-3^{-\alpha})^{-1}$ & \{$\lambda_2$, $\lambda_7$\}\\
$(1+2^{-\alpha}-3^{-\alpha})^{-1}$ & \{$\lambda_{10}$, $\lambda_{15}$\}\\
$-(1-2^{-\alpha}+3^{-\alpha})^{-1}$ & \{$\lambda_3$, $\lambda_6$\}\\
$(1-2^{-\alpha}+3^{-\alpha})^{-1}$ & \{$\lambda_{11}$, $\lambda_{14}$\}\\
$-(1-2^{-\alpha}-3^{-\alpha})^{-1}$ & \{$\lambda_4$, $\lambda_5$\}\\
$(1-2^{-\alpha}-3^{-\alpha})^{-1}$ & \{$\lambda_{12}$, $\lambda_{13}$\}\\
$-2^\alpha$ & \{$\lambda_4$, $\lambda_8$\}\\
$2^\alpha$ & \{$\lambda_9$, $\lambda_{13}$\}\\
$-(2-2^{-\alpha})^{-1}$ & \{$\lambda_6$, $\lambda_8$\}\\
$(2-2^{-\alpha})^{-1}$ & \{$\lambda_9$, $\lambda_{11}$\}\\
\hline
\end{tabular}
\label{depair}
\end{table}
\end{center}

The features of $\Gamma_\text{tr}$ for even and odd $N$ with $\alpha=0$ as shown in Fig.~\ref{ai} can be understood by inspecting their Hamiltonian matrices. Let us analyze the case $N=2$ explicitly, whose Hamiltonian matrix takes the form
\begin{equation}
H_{N=2}=\begin{bmatrix}
 -2J+U & -i\sqrt{2}\Gamma&0\\
 -i\sqrt{2}\Gamma & -U&-i\sqrt{2}\Gamma\\
0&-i\sqrt{2}\Gamma & 2J+U 
\end{bmatrix}.\label{hn2}
\end{equation}
When $\Gamma=0$, the three eigenvalues of $H_{N=2}$ are $\lambda_1=-2J+U$, $\lambda_2=-U$ and $\lambda_3=2J+U$; at the points $U=\pm J$, $H_{N=2}$ has a two-fold degenerate eigenvalue ($J\neq0$), and infinitesimal $\Gamma$ is capable of triggerring the $\mathcal {PT}$-symmetry-breaking transition as shown in Fig.~(\ref{ai}) at $U=\pm J$ with $\Gamma_\text{tr}=0$. 
In another limiting situation $|U/J|\gg1$, the Rabi term $J$ can be neglected with respect to $U$ in the diagonal elements of Eq.~(\ref{hn2}), and the eigenvalues $\lambda$ of $H_{N=2}$ are determined by the secular equation, $(U-\lambda)[\lambda^2-U^2+4\Gamma^2]=0$; the coalescence is found to occur at $\Gamma_{\rm tr}=|U|/2$, resulting in the asymptotically linearly divergent boundary shown in Fig.~\ref{ai}. 

To understand the difference between $N$ being odd and even, Let us also look at the case $N= 3$. The Hamiltonian takes the form
\begin{equation}
H_{N=3}=\begin{bmatrix}
-3J+3U  & -i\sqrt{3}\Gamma & 0 & 0\\
 -i\sqrt{3}\Gamma & -J-U & -i2\Gamma & 0\\
 0 & -i2\Gamma & J-U & -i\sqrt{3}\Gamma\\
 0 & 0 & -i\sqrt{3}\Gamma &3J+3U
\end{bmatrix}.
\label{hn3}
\end{equation}
When $\Gamma=0$, the eigenenergies of Eq.~(\ref{hn3}) are $\lambda_1=-3J+3U, \lambda_2=-J-U,\lambda_3=J-U, \lambda_4=3J+3U$; for nonzero $J$, at the points $U/J=\pm1/2$, either the pair $\{\lambda_1,\lambda_2\}$ or the pair $\{\lambda_3,\lambda_4\}$ become degenerate. These degeneracies give rise to the zeros of $\Gamma_{\rm tr}/J$ for $N=3$ in Fig.~\ref{ai}. 
In the limit $|U/J|\gg1$, we identify a particular subspace spaned by $|3/2,1/2\rangle$ and $|3/2,-1/2\rangle$ in which the diagonal elements of $H_{N=3}$ are approximately the same and the off-diagonal elements are nonzero to first order of $\Gamma$. The Hamiltonian in the subspace is given by
\begin{equation}
\label{Eq.10}
H^{(sub)}_{N=3}=\begin{bmatrix}
 -J-U & -i2\Gamma\\
 -i2\Gamma & J-U
\end{bmatrix},
\end{equation}
whose eigen polynomial is quadratic, i.e.,
\begin{equation}
\label{Eq.11}
f(\lambda)=\lambda^2+2U\lambda-J^2+U^2+4\Gamma^2,
\end{equation}
from which we obtain the discriminant $\Delta=J^2-4\Gamma^2$. Thus coalescing of eigenenergies occurs at $\Gamma_{\rm tr}/J=1/2$ as shown in Fig.~\ref{ai}. 

The above analysis can be generalized to arbitrary $N$. As stated in the main text, the zeros of $\Gamma_{\rm tr}$
are found to located at $J/U=1-2m$ ($m$ cannot take $1/2$ because $J\neq0$), confined in the region $-1\le U/J\le1$. Thus, there are $N$ zeros of $\Gamma_{\rm tr}$ for even $N$, and $N-1$ zeros for odd $N$.
In the large $|U|$ limit, for odd $N=2k+1$, we attribute likewise the transition to the subspace spanned by $|N/2,1/2\rangle$ and $|N/2,-1/2\rangle$; the Hamiltonian matrix in the subspace is
\begin{equation}
\label{Eq.12}
H^{(sub)}_{N=2k+1}=\begin{bmatrix}
 -J-(N-1)U/2 & -i(N+1)\Gamma/2\\
-i(N+1)\Gamma/2 & J-(N-1)U/2
\end{bmatrix}.
\end{equation} 
Coalescing of eigenenergies requires the discriminant $\Delta= \left [ J^2-(N+1)^2\Gamma^2/4\right ] =0$, which yields  $\Gamma_{\rm tr}\to 2J/ (N+1)$ as $|U/J|\to\infty$. 
On the other hand, for even $N=2k$, we expect $\Gamma_{\rm tr}\sim |U|$ in the limit $|U/J|\gg1$ as learned from the $N=2$ case; close to the transition, all the nonzero matrix elements of Eq.~(\ref{hm}) are $\sim U$ such that perturbation treatment is not applicable. Nevertheless, if one truncated the Hamiltonian into the subspace spanned by $|N/2,m\rangle$ with $m=-1,0,1$, one would find the $3\times3$ Hamiltonian matrix in the subspace having the form
\begin{equation}
\label{Eq.10}
H^{(sub)}_{N=2k}=\begin{bmatrix}
-2J-\frac{N-4}{2}U & -i\sqrt{\frac{N^2+2N}{4}}\Gamma & 0\\
-i\sqrt{\frac{N^2+2N}{4}}\Gamma & -\frac{N}{2}U & -i\sqrt{\frac{N^2+2N}{4}}\Gamma\\
0	& -i\sqrt{\frac{N^2+2N}{4}}\Gamma & 2J-\frac{N-4}{2}U 
\end{bmatrix}.
\end{equation} 
Two eigenvalues of $H^{(sub)}_{N=2k}$  coalesce when the discriminant $\Delta\approx\left [4U^2-(2N^2+4N)\Gamma^2\right] =0$, yielding $\Gamma_{\rm tr}\to\left|U\right|/\sqrt{N^2/2+N}$. This truncation into the subspace captures the scaling $\Gamma_{\rm tr}\sim c |U|$ for even $N$ in the limit $|U|/J\to\infty$, and the general trend that the prefactor $c$ decreases with larger $N$; nevertheless to determine $c$ accurately, one needs to take the full Hamiltonian. 

When $\alpha$ becomes nonzero, more zeros of $\Gamma_{\rm tr}$ appear and their locations are not confined in the region $-1\le U/J\le1$ any more, as shown in Figure~\ref{fig:3}. For example, for $N=3$, the Hamiltonian takes form
$
H_{\alpha,N=3}  =H_{N=3} +U\left(\frac1{2^\alpha}-1\right)\sigma_x^1\sigma_x^3\label{finite_ranged_N=3}.\label{ha3}
$
When $\Gamma=0$, since $[\sigma_x^j,H_{\alpha,N=3} ]=0$, we denote the eigenvectors of $H_{\alpha,N=3}$ by $|y\rangle\equiv|x_3\rangle\otimes|x_2\rangle\otimes|x_1\rangle$ where $y=x_3\times 2^2+x_2\times 2+x_1+1$, $\sigma_x^j|x_j\rangle=(-1)^{x_j}|x_j\rangle$, and $x_j=0,1$.
The corresponding eigenvalues are $\lambda_y=J\sum_{j=1}^3 (-1)^{x_j}+U\left[\left(\sum_{j=1}^3 (-1)^{x_j}\right)^2-3\right]/2+U\left(1/2^\alpha-1\right)(-1)^{x_1+x_3}$. As $U/J$ is tuned, some $\lambda_y$ occur to be degenerate. 
Table \ref{depair} lists pairs of degenerate $\{\lambda_y, \lambda_{y'}\}$ for $N=3,4$ at different values of $U/J$, whose corresponding eigen-states have the property $\langle y|\sum_j\sigma_z^j|y'\rangle\neq0$. These degeneracies give rise to the zeros of $\Gamma_{\rm tr}$ in Fig.~\ref{fig:3}. The scaling of $\Gamma_{\rm tr}$ for large $|U/J|$ is still $\sim |U|^1$ for even $N$ and $\sim |U|^0$ for odd $N$; which is independent of $\alpha$.

\end{document}